\documentclass[12pt]{article}
\usepackage[top=1.in, bottom=1.in, left=1in, right=1in]{geometry}
\usepackage{caption}
\usepackage{epsfig}
\usepackage{color}
\usepackage{psfrag}
\usepackage{verbatim}
\usepackage{amsopn}
\usepackage{appendix}
\usepackage{dsfont,amsmath,amssymb,color,units,pstricks}
\usepackage{amsfonts}
\usepackage{graphicx}
\usepackage{lscape}
\usepackage{tikz}
\usepackage{hyperref}
\usetikzlibrary{matrix}

\newcommand\be{\begin{equation}}
\newcommand\bi{\begin{itemize}}
\newcommand\bea{\begin{eqnarray} \nonumber }
\newcommand\ee{\end{equation}}
\newcommand\ei{\end{itemize}}
\newcommand\eea{\end{eqnarray}}

\newtheorem{mydef}{Definition}

\begin{document}

\unitlength = 1mm
\title{Why Do Markets Crash ? Bitcoin Data Offers Unprecedented Insights} 
\date{}
\author{ Jonathan Donier$^{1,2,3}$ and Jean-Philippe Bouchaud$^{1,4}$}

    \maketitle
    \noindent\small{$1$: Capital Fund Management, 23-25 Rue de l'Universit\'e, 75007 Paris, France\\}
    \noindent\small{$2$: Laboratoire de Probabilit\'es et Mod\`eles Al\'eatoires, Universit\'e Pierre et Marie Curie (Paris 6), 4 Place Jussieu, 75005 Paris\\}
    \noindent\small{$3$: Ecole des Mines ParisTech, 60 Boulevard Saint-Michel, 75006 Paris\\}
    \noindent\small{$4$: CFM-Imperial Institute of Quantitative Finance, Department of Mathematics, Imperial College, 180 Queen's Gate, London SW7 2RH}
\normalsize

\begin{abstract}
Crashes have fascinated and baffled many canny observers of financial markets. In the strict orthodoxy of the efficient market theory, crashes must be due to sudden changes of the fundamental valuation of assets. However, 
detailed empirical studies suggest that large price jumps cannot be explained by news and are the result of endogenous feedback loops. Although plausible, a clear-cut empirical evidence for such a scenario is still lacking. 
Here we show how crashes are conditioned by the market liquidity, for which we propose a new measure inspired by recent theories of market impact and based on readily available, public information. Our results open the possibility 
of a dynamical evaluation of liquidity risk and early warning signs of market instabilities, and could lead to a quantitative description of the mechanisms leading to market crashes.
\end{abstract}

\tableofcontents

\clearpage

\section{Introduction}\label{sec:crash}

Why do market prices move? This simple question has fuelled fifty years of academic debate, reaching a climax with the 2013 Nobel prize in economics, split between Fama and Shiller who promote radically different views on the question~\cite{shiller2013sharing}. Whereas Fama argues that markets are efficient and prices faithfully reflect fundamental values, Shiller has shown that prices fluctuate much more than what efficient market theory would suggest, and 
has insisted on the role of behavioural biases as a source of excess volatility and price anomalies. 
Of particular importance is the origin of the largest changes in prices, aka market crashes, that may have dire consequences not 
only for market participants but also for the society as a whole~\cite{taleb2010black}. It is fair to say that after centuries of market folly~\cite{mackay2012extraordinary,kindleberger2011manias,sornette2009stock,reinhart2009time}, there is no 
consensus on this issue. Many studies~\cite{fair2000events, joulin2008stock, cornell2013moves} have confirmed the insight of Cutler, 
Poterba \& Summers~\cite{cutler1989moves} who concluded 
that {\it  [t]he evidence that large market moves occur on days without identifiable major news casts doubts on the view that price movements are fully explicable by news...}. The fact that markets appear to crash in the absence of any remarkable event suggests that destabilising feedback loops of behavioural origin may be at play~\cite{smith1988bubbles, lillo2005key, hommes2005coordination, bouchaud2013crises}. Although plausible, a clear-cut empirical evidence for such an endogenous scenario is still lacking. After all, crashes are not that frequent and a convincing statistical analysis is difficult, 
in particular because of the lack of relevant data about the dynamics of supply and demand during these episodes.

In this respect, the Bitcoin~\cite{nakamoto2008bitcoin,ali2014economics,bohme2014bitcoin} market is quite unique on many counts. In particular, the absence of any compelling way to assess the fundamental price of Bitcoins makes the behavioral hypothesis highly plausible. For our purpose, the availability of the full order book\footnote{The order book is the record of all intentions to buy or sell at a given point in time, each volume coming with an offering price.} at all times provides precious insights, in particular before and during extreme events. Indeed, at variance with most financial markets where participants hide their intentions, the orders are placed long in advance by Bitcoin traders over large price ranges. Using two highly informative data-sets -- the trade-by-trade MtGox data between December 2011 and January 2014, and the full order book data over the same period -- we analyse in depth the liquidity of the Bitcoin market. We find that what caused the crash was not the selling pressure per se, but rather the dearth of buyers that stoked the panic. Following up on this observation, we show that three different liquidity measures that aim at quantifying the presence of buyers (or sellers) are highly correlated and correctly predict the amplitude of potential crashes. Whereas two of them are direct probes of the prevailing liquidity but difficult to access on financial markets, the third one -- which is also firmly anchored theoretically~\cite{donier2014fully} -- only uses readily available, public information on traded volumes and volatility, and is therefore a promising candidate for monitoring the propensity of a market to crash. 

\section{Anatomy of April 10, 2013 crash}\label{sec:crash}

Amongst all crashes that happened on the Bitcoin and for which we found some data, the April 10, 2013 crash is probably the most interesting one since on that day the price dropped by more than $50\%$ of its value in a few hours. At that time, MtGox was by far the leading exchange (its market share was over $80\%$ on the BTC/USD spot market) so our data-set captures a large fraction of the investors' behaviour. Intuitively, the main driver of market crashes is the mismatch between the aggregate market order flow imbalance (${\cal O}$, defined below) that becomes strongly negative and the prevailing liquidity on the buy side, {i.e. the density of potential buyers below the current price}. Whereas the former quantity can be easily reconstructed from the series of trades, the notion of ``prevailing liquidity'' is only at best ambiguous. It is only when the price starts heading down, that one expects most of the interested buyers to declare themselves and post orders in the order book. Therefore, the liquidity cannot in principle be directly inferred from the information on the publicly available order book. The dynamic nature of liquidity has been clearly evidenced~\cite{weber2005order,bouchaud2006random}, and has led to the notion of ``latent'' liquidity that underpins recent theories of impact in financial markets~\cite{toth,mastromatteo2014agent,mastromatteo2014anomalous,donier2014fully}.

\begin{figure}[ht!]
\centering
\includegraphics[width=0.8\textwidth]{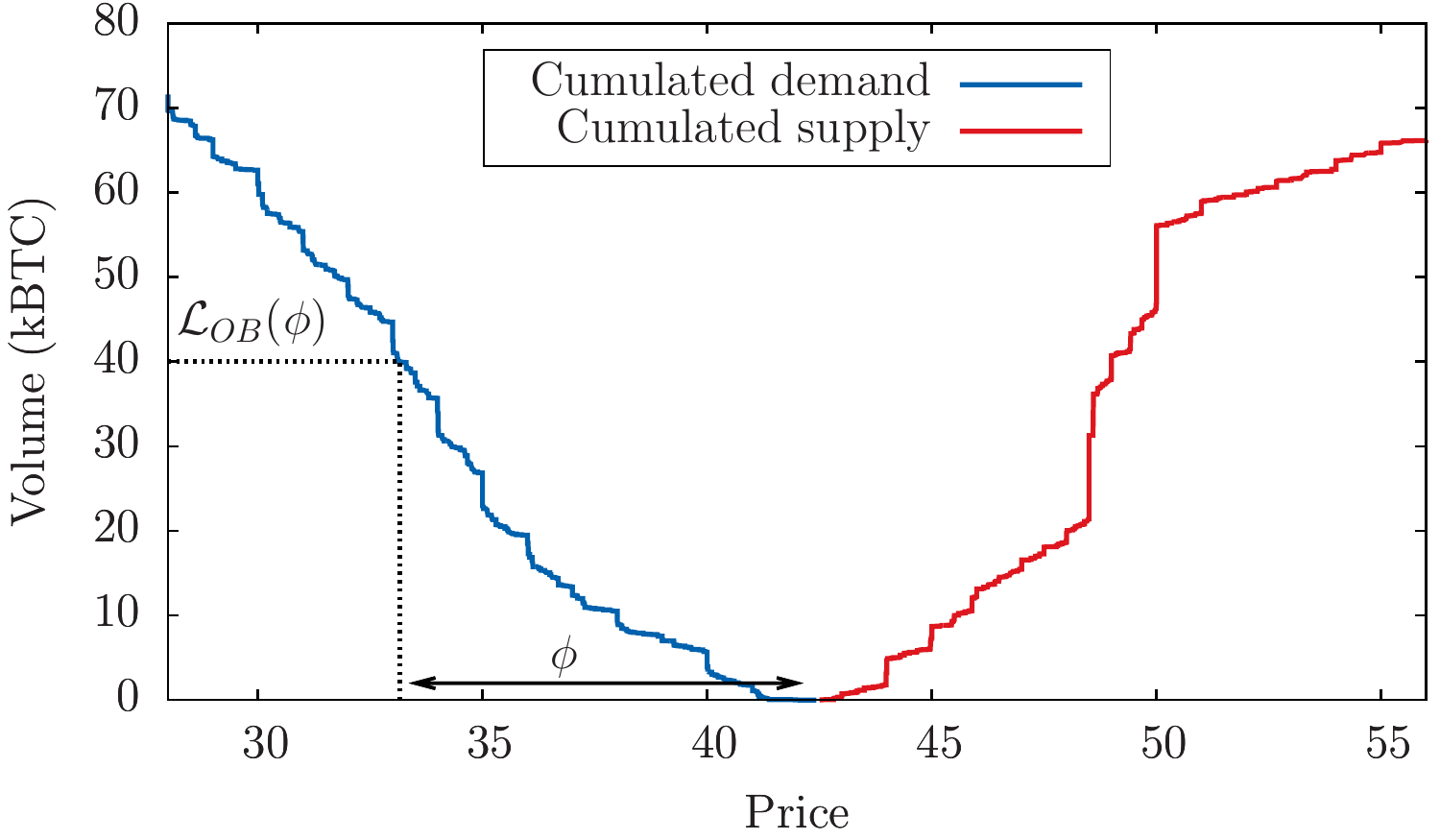}
\caption{ {\bf Instantaneous cumulated order book.} Snapshot of the cumulated supply and demand displayed on the order book {taken on March 8, 2013}, with a graphical representation of the order book liquidity $\mathcal{L}_{OB}(\phi)$ defined in Def.~\ref{def:liq3}.}
\label{fig:OB}
\end{figure}

\begin{figure}[ht!]
\centering
\includegraphics[width=0.8\textwidth]{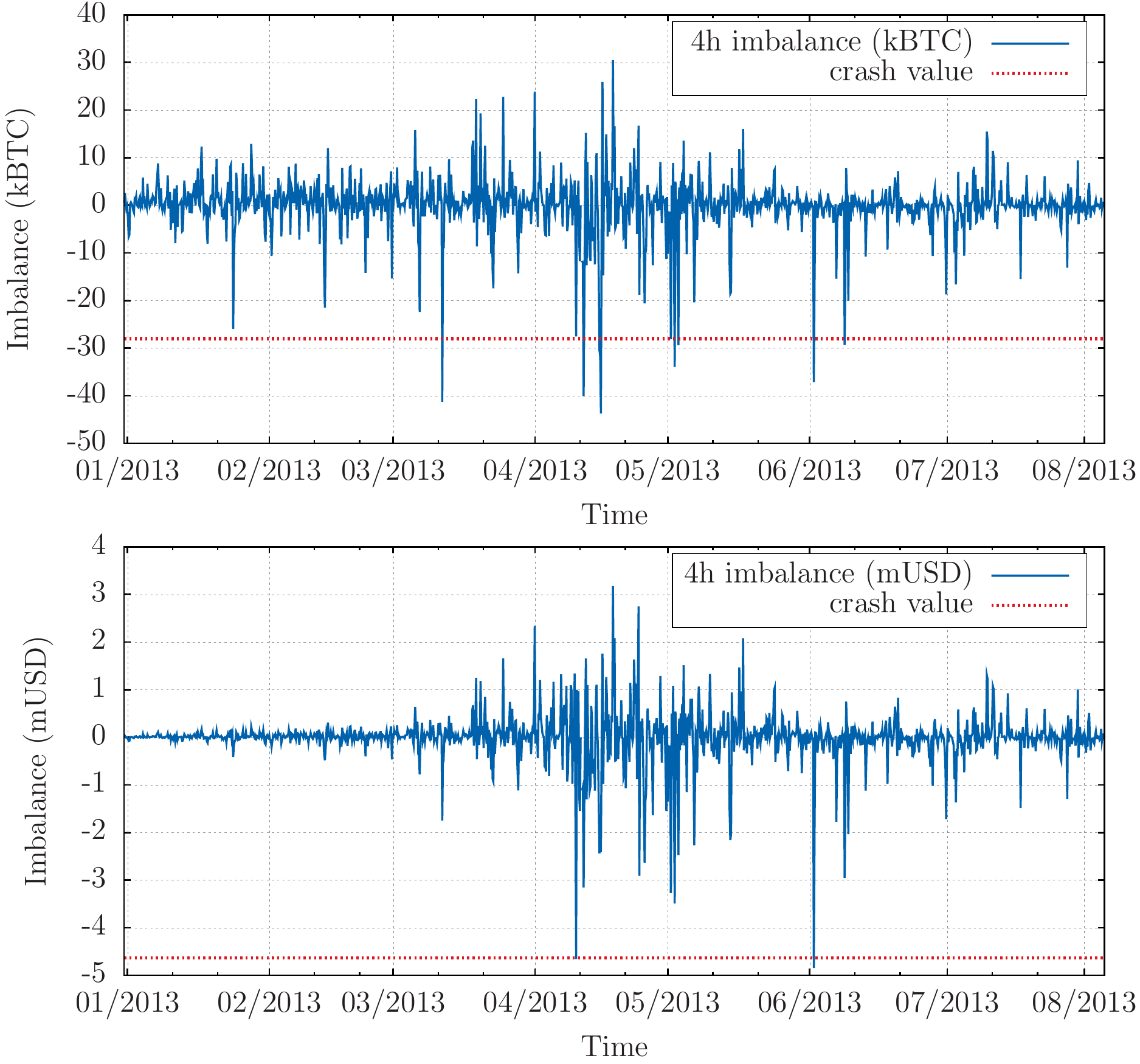}
\caption{ {\bf Order flow imbalances in USD and BTC.} Top:  Aggressive imbalance in order flow $\sum_i \epsilon_i v_i$ (where $\epsilon_i=\pm 1$ is the sign of the transaction, and $v_i$ its volume in Bitcoins), aggregated over periods of 4 hours between January 2013 and August 2013, expressed in Bitcoins. April 10, 2013 (for which the realised imbalance is represented as a dashed horizontal line) does not appear as an outlier. Bottom: aggressive imbalance in order flow $\sum_i \epsilon_i v_i p_i$, aggregated over periods of 4 hours between January 2013 and August 2013 and expressed in USD. April 10, 2013 now clearly appears as an outlier.}
\label{fig:sellers_BTC}
\end{figure}

However, Bitcoin is quite an exceptional market in this respect, since a large fraction of the liquidity is not latent, but actually posted in the
order book -- possibly resulting from less strategic participants on a still exotic market -- and thus directly observable (see Fig.~\ref{fig:OB}). A more quantitative analysis indeed shows that typically $30-40\%$ of the volume traded during the day is already present in the order book in the morning. This is to be compared with a ratio below $1\%$ on more traditional financial markets, say stocks \footnote{The total volume in the order book 
of major stocks is 5-10 times the volume at the best quotes, which is itself $\sim 10^{-3}$ of the daily turnover, see e.g. \cite{wyart2008relation}.}. This allows us to test in detail the respective roles of aggregate imbalance and 
liquidity in the triggering of market crashes. We first study the ``aggressive''  order flow defined as the aggregated imbalance of market orders for every 4 hours window between January 2013 and August 2013. In fact, two definitions are possible. 
One is defined as the average of the signed number of Bitcoin contracts sent as market orders\footnote{A market order is an order to trade immediately at the best available price.
Because of this need for immediacy, one often refers to them as aggressive orders.} ${\cal O}_B = \sum_i \epsilon_i q_i,$ where each $i$ is a different market order of sign $\epsilon_i$ ($\epsilon_i=+1$ for buyer-initiated trades and $-1$ for seller-initiated trades) and 
number of contracts $q_i$, and the sum runs over consecutive trades in a 4 hour window. The second is the volume imbalance expressed in USD: ${\cal O}_{\$} = \sum_i \epsilon_i q_i p_i,$ where $p_i$ is the 
$i$-th transaction price. These two quantities are shown in Fig.~\ref{fig:sellers_BTC} and reveal that (a) large sell episodes are more intense than large buy episodes; 
(b) when expressed in Bitcoin, the sell-off that occurred on April, 10 (of order of 30,000 BTC {on a 4h window}) is not more spectacular than several other sell-offs that happened before or after that day; 
(c) however, when expressed in USD, the April 10 sell-off indeed appears as an outlier. 

\begin{figure}[ht!]
\centering
\includegraphics[width=0.85\textwidth]{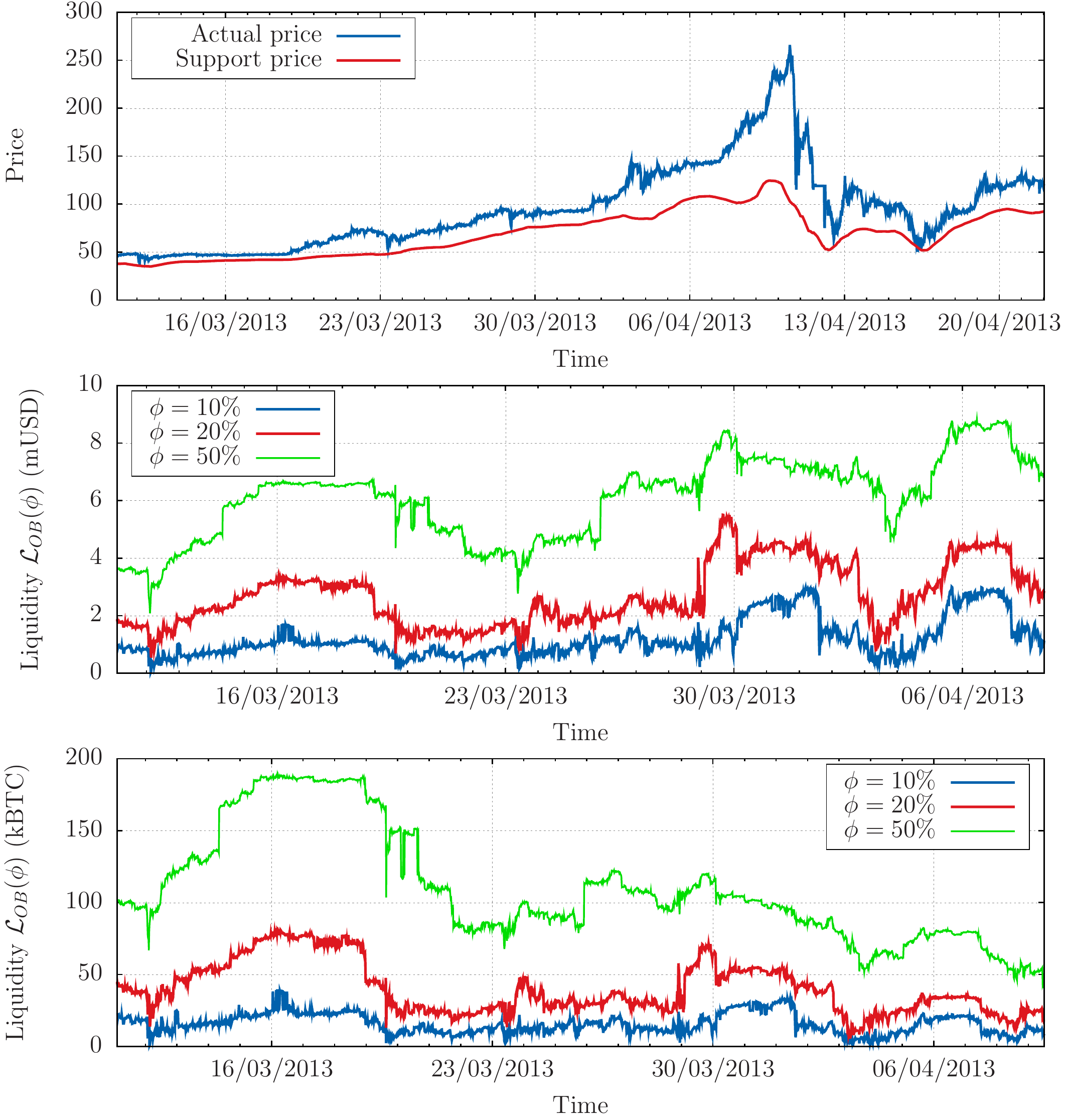}
\caption{ {\bf Liquidity and support price.} Top: Actual price $p_t$ (blue) vs. support price $p_S^{\text{40k}}$ (red) defined as the price that would be reached if a typical sell-off of 40,000 BTC was to occur instantaneously. Note that $p_S^{\text{40k}}$ is $\approx 50 \%$ below the price $p_t$ just before the crash, explaining the 
order of magnitude of the move that happened that day. Middle (resp. Bottom): Buy volume ${\cal L}_{\text{OB}}(\phi)$ in USD (resp. BTC) in the order book, during the months preceding the crash of April 10, 2013, measured as the volume between the current price $p_t$ and $p_t(1-\phi)$ where $\phi=10\%$, $20\%$ and $50\%$. One can see that for any quantile the liquidity in USD tended to increase by an overall factor $\simeq 2$ during the period, while the liquidity in BTC was decreased by a factor $\simeq 2-3$ as an immediate consequence of the bubble.}
\label{fig:crash_bande}
\end{figure}

The difference between ${\cal O}_B$ and ${\cal O}_{\$}$ originates from the fact that a large fraction of this selling activity occurred at the peak of the ``bubble'' that preceded the crash, see Fig.~\ref{fig:crash_bande}, top. The BTC price rose from $\$13$ in early January to $\$260$ just before the crash. In Fig.~\ref{fig:crash_bande}, we represent a ``support'' level $p_S^{\text{40k}}$ such that the total quantity of buy orders between $p_S^{\text{40k}}$ and the current price $p_t$ is 40,000 BTC, see Fig.~\ref{fig:OB}. One notices that the price dramatically departed from the support price during the pre-crash period, which is a clear sign that Bitcoin price was engaged in a bubble. Although the liquidity expressed in USD was actually {\it increasing} during that period (see Fig.~\ref{fig:crash_bande}, middle), the BTC price increased even faster, resulting in a  thinner and thinner 
liquidity on the buy side of the order book {\it expressed in BTC}, see  Fig.~\ref{fig:crash_bande}, bottom.  This  
scenario is precisely realised in some Agent Based Models of markets~\cite{giardina2003bubbles}.

The conclusion of the above analysis, that may appear trivial, is that the crash occurred because the price was too high, and buyers too scarce to resist 
the pressure of a sell-off. 
More interesting is the fact that the knowledge of the volume present in the 
order book allows one to estimate an expected price drop of $\approx 50 \%$  in the event of a large -- albeit not extreme -- sell-off. Of course, the 
possibility to observe the full demand curve (or a good approximation thereof) is special to the Bitcoin market, and not available in more mainstream markets { where the publicly displayed liquidity is only of order $1\%$ 
of the total daily traded volume}. Still, as we show now, one can built accurate proxies of the latent liquidity using observable quantities only, opening the path to early warning signs of an impeding crash.

\section{Three definitions of ``liquidity''}\label{sec:prediction}

More formally, the market liquidity measure discussed above is defined as:

\begin{mydef}\label{def:liq3}
The \emph{order-book liquidity} ${\cal L}_{\text{OB}}(\phi)$ (on the buy side) is such that (cf. Fig.~\ref{fig:OB} above):
\be
\int_{p_t(1-\phi)}^{p_t}{\rm d}p \rho(p,t):={\cal L}_{\text{OB}}(\phi) \;,
\ee
(and similarly for the sell-side). In the above equation, $p_t$ is the price at time $t$ and $\rho(p,t)$ is the density of demand that is materialised on the order book at price $p$ and at time $t$. 

Conversely, the price drop $-\phi^* p_t$ expected if a large instantaneous sell-off of size $Q^*$ occurs is such that:
\be
\phi^* = {\cal L}_{\text{OB}}^{-1}( Q^*),
\ee
where ${\cal L}_{\text{OB}}^{-1}$ is a measure of illiquidity.
\end{mydef}

\begin{figure}[ht!]
\centering
\includegraphics[width=.9\textwidth]{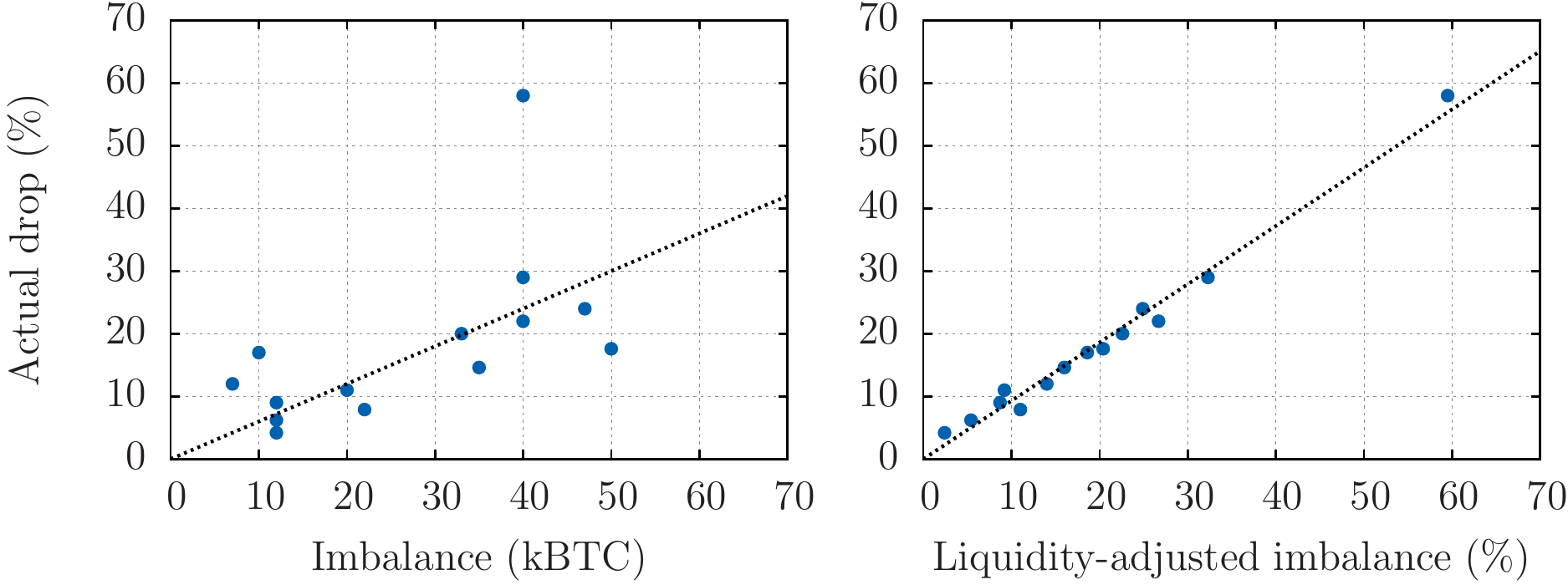}
\caption{ {\bf Forecast of crashes amplitudes using order book volumes.} For the 14 most extreme negative returns 
that have occurred between Jan 1, 2013 and Apr 10, 2013, we compare the realised return with: (Left) the net imbalance ${\cal O}_B$ during the period 
(usually a few hours) and (Right) the \emph{liquidity-adjusted} imbalance ${\cal L}_{\text{OB}}^{-1}({\cal O}_B)$. 
This illustrates the relevance of the ${\cal L}_{\text{OB}}$ liquidity measure to predict the amplitudes of crashes -- even in the most extreme cases.}
\label{fig:main_drops}
\end{figure}

An \textit{a posteriori} comparison between realised returns and the liquidity-adjusted imbalance for the 14 most extreme negative returns
that have occurred between Jan 1, 2013 and Apr 10, 2013 is shown in Fig.~\ref{fig:main_drops}. {These events, which corresponds to dramatic jumps in the cumulated order flow process, are found to have a characteristic scale of about 4h with a standard deviation of 2.5h, justifying the choice made in Fig.~\ref{fig:sellers_BTC} to plot imbalances at a 4h time scale.} The analysis shows 
that the quantity ${\cal L}_{\text{OB}}^{-1}({\cal O}_B)$ nearly perfectly matches crashes amplitudes, vindicating the hypothesis that 
most of the liquidity is indeed present in the visible order book for the Bitcoin.

However, as recalled above, the visible order book on standard financial markets usually contains a minute fraction of the real intentions 
of the agents. Therefore the use of ${\cal L}_{\text{OB}}(\phi)$ deduced from the observable order book would lead to a tremendous underestimation of the 
liquidity in these markets~\cite{sandaas2001adverse, weber2005order}. Liquidity is in fact a dynamic notion, that {reveals itself progressively as} a reaction 
(possibly with some lag) to the incoming order flow~\cite{weber2005order,bouchaud2006random}. Another definition of liquidity, 
that accounts for the progressive appearance of the latent liquidity as orders are executed, is based on a measure of {\it market impact}. 
With enough statistics, the average (relative) price move $I(Q)=\langle \Delta p/p \rangle$ induced by the execution of a meta-order\footnote{A meta-order is 
a sequence of individual trades generated by the same trading decision but spread out in time, so as to get a better price and/or not to be detected~\cite{toth}.} 
can be measured as a function of their total volume $Q$. Since these meta-orders are executed on rather long time scales (compared to the transaction frequency), it is reasonable to think that their impact reveals the ``true'' latent liquidity of markets~\cite{toth,mastromatteo2014agent,mastromatteo2014anomalous,donier2014fully}. This leads us to a second definition of liquidity, based on market impact:

\begin{mydef}\label{def:liq1}
The \emph{impact liquidity} ${\cal L}_{\text{I}}(\phi)$ is defined as the volume of a meta-order that moves, on average, the price $p_t$ by $\pm \phi p_t$, or, more precisely, ${\cal L}_{\text{I}}(\phi)$ is fixed by the condition:
\be
I({\cal L}_{\text{I}}(\phi)) = \phi,
\ee
since the impact $I(Q)$ is usually measured in relative terms. As above, the price drop expected if a large sell-off of volume imbalance $Q^*$ occurs is
simply given by ${\cal L}_{\text{I}}^{-1}(Q^*)= I(Q^*)$. 
\end{mydef}

The problem with this second definition is that it requires proprietary data with sufficient statistics, available only to brokerage firms or to active asset managers/hedge funds. It turns out to be also available for Bitcoin~\cite{donier2014million} -- see below. However, a very large number of empirical studies in the last 15 years have established that the impact of meta-orders follows an extremely robust ``square-root law''~\cite{Barra:1997,almgren2005direct,moro2009market, donier2014million, bladon2012agent,toth,kyle2012large,bershova2013non, gomes2014market, mastromatteo2014agent,
brokmann2014slow}. Namely, {\it independently} of the asset class, time period, style of trading and micro-structure peculiarities, one has:
\be\label{eq:impact_emp}
I_{\text{TH}}(Q) \,\, \approx \,\, Y \sigma_\text{d} \sqrt{\frac{Q}{V_\text{d}}} \;,
\ee
where $Y$ is an a-dimensional constant of order unity, $V_\text{d}$ is the daily traded volume and $\sigma_\text{d}$ is the daily volatility. This 
square-root law has now been justified theoretically by several authors, building upon the notion of latent liquidity~\cite{toth,mastromatteo2014agent,mastromatteo2014anomalous,donier2014fully} (see Ref.~\cite{farmer2013efficiency} for an alternative story). 
Assuming that the above functional shape of market impact is correct leads to a third definition of liquidity:
\begin{mydef}\label{def:liq2}
The \emph{theoretical liquidity} ${\cal L}_{\text{TH}}(\phi)$ is the theoretical volume of a meta-order required to move the price $p_t$ by $\pm \phi p_t$
according to formula Eq.~(\ref{eq:impact_emp}) above, i.e.:
\be
I_{\text{TH}}({\cal L}_{\text{TH}}(\phi)) = \phi.
\ee
\end{mydef}
Together with Eq.~(\ref{eq:impact_emp}), this amounts to consider $\sigma_\text{d}/\sqrt{V_d}$ as a measure of market illiquidity. 
Clearly, since both $\sigma_\text{d}$ and ${V_d}$ can be estimated from public market data, this last 
definition of liquidity is quite congenial. It was proposed in Ref.~\cite{caccioli2012impact} as a proxy to obtain impact-adjusted marked-to-market 
valuation of large portfolios, and tested in Ref.~\cite{kyle2012large} on five stock market crashes, with very promising results. However, 
there is quite a leap of faith in assuming that our above three definitions are -- at least approximately -- equivalent. 
This is why the Bitcoin data is quite unique since it allows one to measure all three liquidities 
${\cal L}_{\text{OB}}, {\cal L}_{\text{I}},{\cal L}_{\text{TH}}$ and test quantitatively that they do indeed reveal the very same
information.

\section{Comparing the liquidity measures}

We measured the order book liquidity ${\cal L}_{\text{OB}}$ {at the daily scale by averaging the volume present at all prices in the buy side of the order book for each day}. The empirical impact is obtained following Ref.~\cite{donier2014million} by measuring the full $I(Q)$, obtained as an average over all meta-orders of a given volume $Q$ on a given day. Finally, the theoretical impact Eq.~(\ref{eq:impact_emp}) is obtained by measuring both the traded volume of the
day $V_\text{d}$ and the corresponding volatility $\sigma_\text{d}$ \footnote{defined as  $\sigma_\text{d}^2 = \frac{1}{T}\sum_{t=1}^T\left ( 0.5 \text{ln} \left ( H_t/L_t\right )^2 - \left (2\text{ln}(2)-1\right ) \text{ln} \left ( C_t/O_t\right )^2 \right )$ where $O_t$/$H_t$/$L_t$/$C_t$ are the open/high/low/close prices of the sub-periods~\cite{garman1980estimation}.}. {The daily scale has been chosen so as to average out market noise and intraday patterns in the measure of ${\cal L}_{\text{I}}^{-1}$ and ${\cal L}_{\text{TH}}^{-1}$, while remaining reactive to liquidity fluctuations: Fig.~\ref{fig:crash_bande} indeed shows how much liquidity can fluctuate in a few days.}

\begin{figure}[ht!]
\centering
\includegraphics[width=0.85\textwidth]{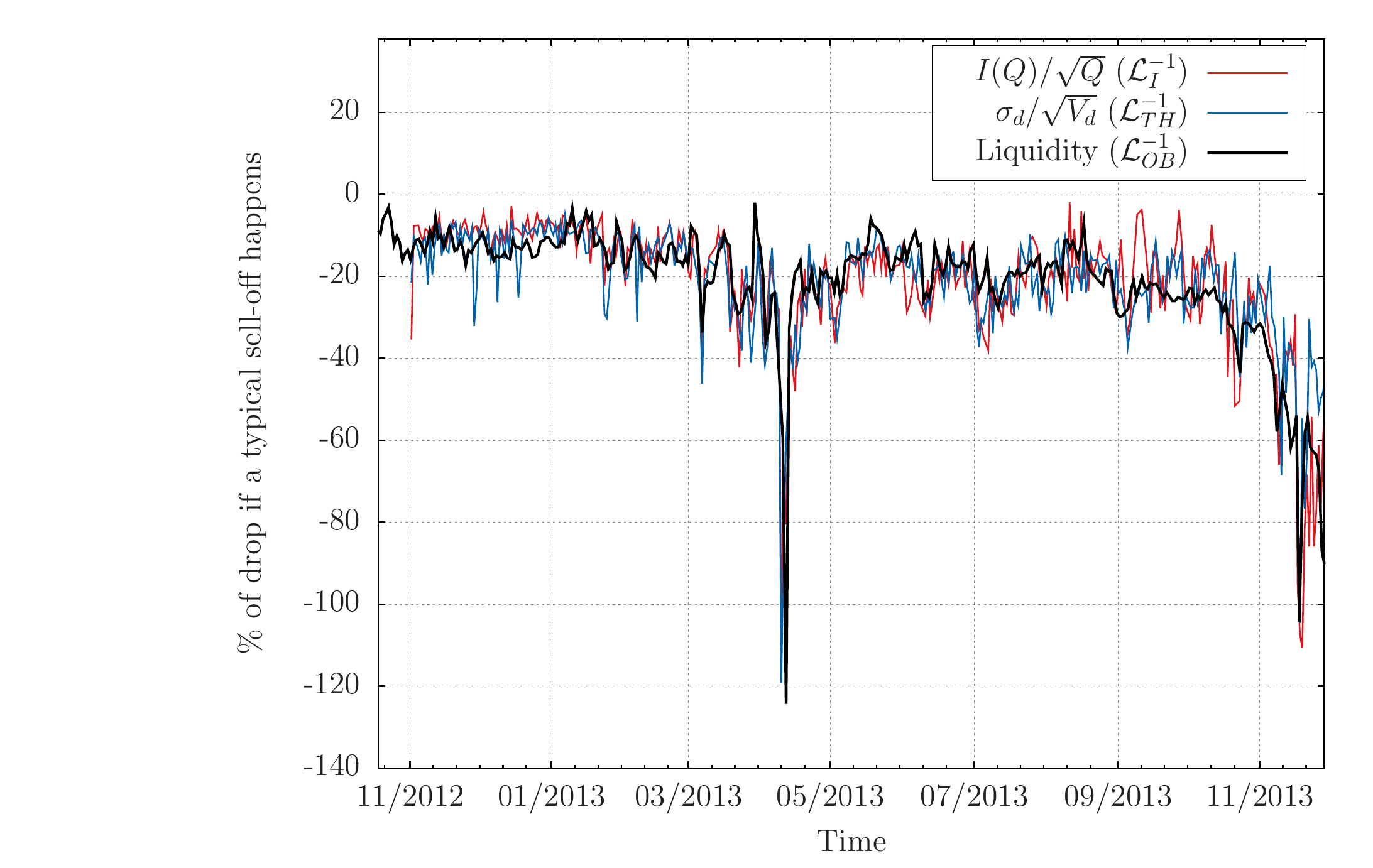}
\caption{ {\bf Comparison between the three (il-)iquidity measures.} Parallel evolution of the three price drops $\phi^*$ deduced from our three estimates of illiquidity  
${\cal L}_{\text{OB}}^{-1}, {\cal L}_{\text{I}}^{-1},{\cal L}_{\text{TH}}^{-1}$ defined above. The estimates based on ${\cal L}_{\text{I}}^{-1},{\cal L}_{\text{TH}}^{-1}$
have been rescaled by a factor $6.10^4$ to match the average order book data prediction.}
\label{fig:liquidity}
\end{figure}

\begin{figure}[ht!]
\centering
\includegraphics[width=0.8\textwidth]{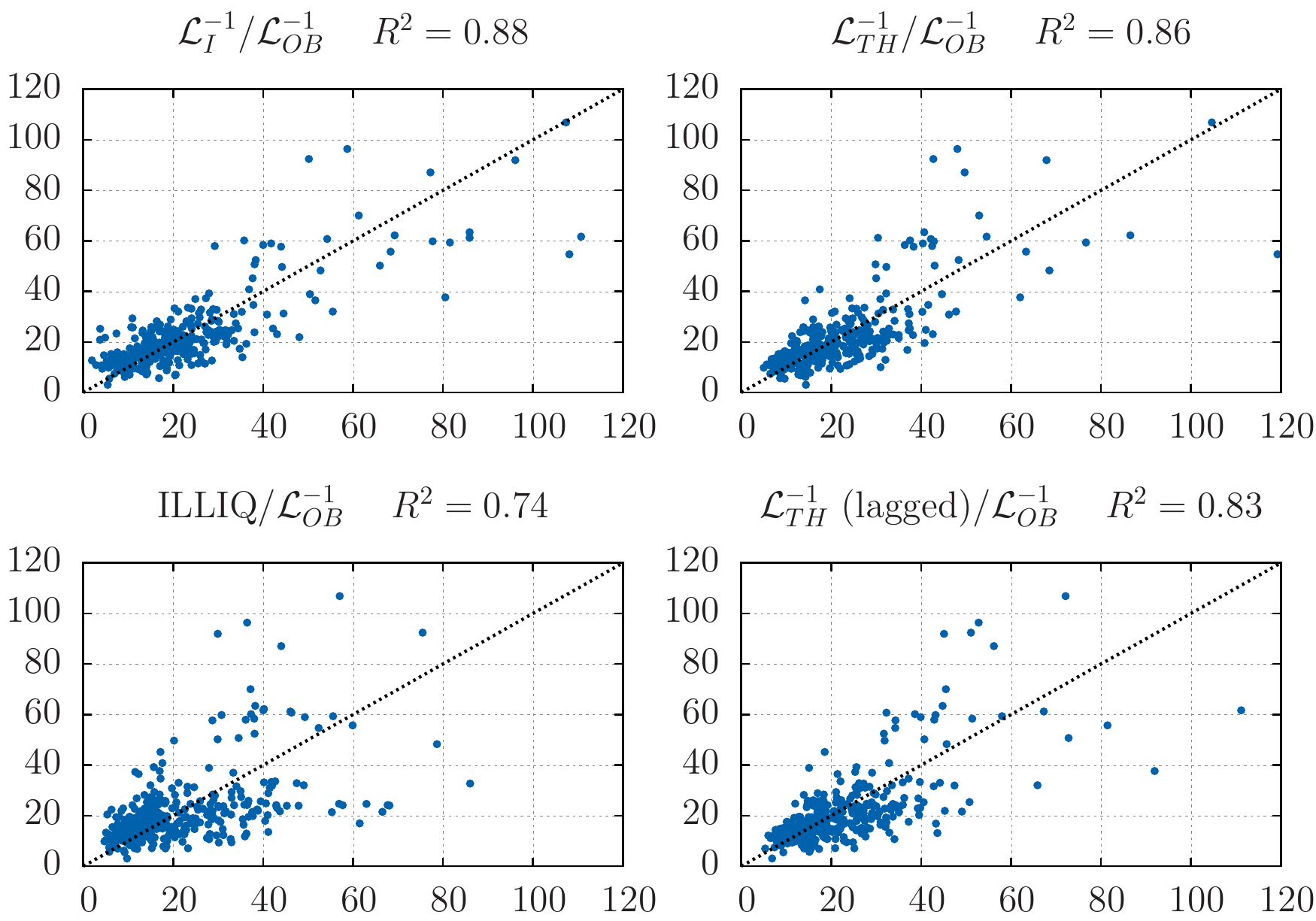}
\caption{ {\bf Regression of the actual (il-)liquidity against the different (il-)liquidity measures.} Regression of the actual illiquidity $\mathcal{L}_{OB}^{-1}$ on three same-day illiquidity measures (after rescaling so that the samples means coincide): The direct measure of orders market impact ${\cal L}_{\text{I}}^{-1}$, the publicly available measure ${\cal L}_{\text{TH}}^{-1}$ that corresponds to the theoretical and empirical impact, and the well-known Amihud {\small{ILLIQ}} measure \protect{~\cite{amihud2002illiquidity}}. Both ${\cal L}_{\text{I}}^{-1}$ and ${\cal L}_{\text{TH}}^{-1}$ outperform {\small{ILLIQ}} ($R^2 \approx 0.86$ vs. $0.74$). 
Note that a high predictability remains when lagging ${\cal L}_{\text{TH}}^{-1}$ by one day ($R^2 \approx 0.83$ vs. $0.71$). The 
regression slopes for the four graphs are, respectively:  $0.9, 0.95, 0.87$ and $0.93$.}
\label{fig:yratio_pred}
\end{figure}

These three estimates allow us to compare, as a function
of time (between November 2012 and November 2013) the expected price drop for a large sell meta-order of size -- say -- $Q^*=40,000$ BTC, see Fig.~\ref{fig:liquidity}. We have rescaled by a constant factor the predictions based on  ${\cal L}_{\text{I}}$ and ${\cal L}_{\text{TH}}$, so as to 
match the average levels. The agreement is quite striking, and shown in a different way in Fig.~\ref{fig:yratio_pred}
as a scatter plot of ${\cal L}_{\text{OB}}^{-1}$ vs ${\cal L}_{\text{I}}^{-1}$ or ${\cal L}_{\text{TH}}^{-1}$, either on the same day, or with a one day lag. As coinciding times, the $R^2$ of the regressions are $\approx 0.86$ and only fall to $\approx 0.83$ with a day lag, meaning that one can use past data
to predict the liquidity of tomorrow. {As a comparison, when using instead Amihud's~\cite{amihud2002illiquidity} measure of illiquidity $\sigma_\text{d}/V_\text{d}$, one obtains $R^2$ of resp. 0.74 and 0.71.}

That the estimates based on ${\cal L}_{\text{I}}$ and ${\cal L}_{\text{TH}}$ match is no surprise since the square-root law was already tested with a high degree of precision on the Bitcoin~\cite{donier2014million}. But that the theoretical measure of liquidity ${\cal L}_{\text{TH}}$ based on easily accessible market data is 
able to track so closely the information present in the whole order book is truly remarkable, and suggests that one can indeed faithfully use ${\cal L}_{\text{TH}}$
on markets where reliable information on the latent order book is absent (as is the case for most markets).

\section{Discussion -- Anticipating crashes?}

Thanks to the unique features of the Bitcoin market, we have been able to investigate some of the factors that determine the propensity of 
a market to crash. Two main features emerge from our study. First, the price level should lie within a range where the underlying demand (resp. supply) is able to support large -- but expected -- 
fluctuations in supply (resp. demand). When the price is clearly out of bounds (for example the pre-April 2013 period for Bitcoin) the market is unambiguously in
a precarious state that can be called a \emph{bubble}. Our main result allows one to make the above idea meaningful in practice. 
We show that three natural liquidity measures (based, respectively, on the knowledge of the full order book, on the average impact of meta-orders, and on the ratio of the volatility to the square-root of the traded volume, $\sigma_\text{d}/\sqrt{V_\text{d}}$) are {\it highly correlated} and do predict the amplitude of a putative crash induced by a given (large) sell order imbalance.

Since the latter measure is entirely based on readily available public information, our result is quite remarkable. It opens the path to a better understanding of crash mechanisms and possibly to early warning signs of market instabilities. However, while we claim that the amplitude of a potential crash can be anticipated, we are of course not able to predict when this crash will happen -- if it happens at all. Still, our analysis motivates better dynamical risk evaluations (like value-at-risk), impact adjusted marked-to-market accounting~\cite{caccioli2012impact} or liquidity-sensitive option valuation models. As a next step, a comprehensive study of the correlation between the realised crash probability and $\sigma_\text{d}/\sqrt{V_\text{d}}$ on a wider universe of stocks -- expanding the work of Ref.~\cite{kyle2012large} -- would be a highly valuable validation of the ideas discussed here.

\normalsize
\paragraph{Acknowledgements:} 
We thank A. Tilloy for his insights on the Bitcoin and for reading the manuscript; P. Baqu\'{e} for reading the manuscript; and J. Bonart for useful discussions.
Bitcoin trades data are available at http://api.bitcoincharts.com/v1/csv/. Bitcoin order book data have been collected by the authors and are available on request.

\bibliographystyle{abbrv}
\bibliography{nature}

\begin{thebibliography}{10}

\bibitem{ali2014economics}
R.~Ali, J.~Barrdear, R.~Clews, and J.~Southgate.
\newblock The economics of digital currencies.
\newblock {\em Bank of England Quarterly Bulletin}, page~Q3, 2014.

\bibitem{almgren2005direct}
R.~Almgren, C.~Thum, E.~Hauptmann, and H.~Li.
\newblock Direct estimation of equity market impact.
\newblock {\em Risk}, 18:5752, 2005.

\bibitem{amihud2002illiquidity}
Y.~Amihud.
\newblock Illiquidity and stock returns: cross-section and time-series effects.
\newblock {\em Journal of financial markets}, 5(1):31--56, 2002.

\bibitem{bershova2013non}
N.~Bershova and D.~Rakhlin.
\newblock The non-linear market impact of large trades: Evidence from buy-side
  order flow.
\newblock {\em Quantitative Finance}, 13(11):1759--1778, 2013.

\bibitem{bladon2012agent}
A.~Bladon, E.~Moro, and T.~Galla.
\newblock Agent-specific impact of single trades in financial markets.
\newblock {\em Physical Review E}, 85(3):036103, 2012.

\bibitem{bohme2014bitcoin}
R.~B{\"o}hme, N.~Christin, B.~G. Edelman, and T.~Moore.
\newblock Bitcoin.
\newblock {\em Journal of Economic Perspectives, Forthcoming}, pages 15--015,
  2014.

\bibitem{bouchaud2013crises}
J.-P. Bouchaud.
\newblock Crises and collective socio-economic phenomena: simple models and
  challenges.
\newblock {\em Journal of Statistical Physics}, 151(3-4):567--606, 2013.

\bibitem{bouchaud2006random}
J.-P. Bouchaud, J.~Kockelkoren, and M.~Potters.
\newblock Random walks, liquidity molasses and critical response in financial
  markets.
\newblock {\em Quantitative finance}, 6(02):115--123, 2006.

\bibitem{brokmann2014slow}
X.~Brokmann, J.~Kockelkoren, and J.-P. Bouchaud.
\newblock Slow decay of impact in equity markets.
\newblock {\em SSRN}, 2014.
\newblock \url{http://papers.ssrn.com/sol3/papers.cfm?abstract_id=2471528}.

\bibitem{caccioli2012impact}
F.~Caccioli, J.~Bouchaud, and D.~Farmer.
\newblock Impact-adjusted valuation and the criticality of leverage.
\newblock {\em Risk}, 2012.

\bibitem{cornell2013moves}
B.~Cornell.
\newblock What moves stock prices: Another look.
\newblock {\em The Journal of Portfolio Management}, 39(3):32--38, 2013.

\bibitem{cutler1989moves}
D.~Cutler, J.~Poterba, and L.~Summers.
\newblock What moves stock prices?
\newblock {\em The Journal of Portfolio Management}, 15(3):4--12, 1989.

\bibitem{donier2014million}
J.~Donier and J.~Bonart.
\newblock A million metaorder analysis of market impact on the bitcoin.
\newblock {\em SSRN}, 2014.
\newblock \url{http://papers.ssrn.com/sol3/papers.cfm?abstract_id=2536001}.

\bibitem{donier2014fully}
J.~Donier, J.~Bonart, I.~Mastromatteo, and J.-P. Bouchaud.
\newblock A fully consistent, minimal model for non-linear market impact.
\newblock {\em SSRN}, 2014.
\newblock \url{http://papers.ssrn.com/sol3/papers.cfm?abstract_id=2531917}.

\bibitem{fair2000events}
R.~Fair.
\newblock Events that shook the market.
\newblock 2000.

\bibitem{farmer2013efficiency}
J.~D. Farmer, A.~Gerig, F.~Lillo, and H.~Waelbroeck.
\newblock How efficiency shapes market impact.
\newblock {\em Quantitative Finance}, 13(11):1743--1758, 2013.

\bibitem{garman1980estimation}
M.~Garman and M.~Klass.
\newblock On the estimation of security price volatilities from historical
  data.
\newblock {\em Journal of business}, pages 67--78, 1980.

\bibitem{giardina2003bubbles}
I.~Giardina and J.-P. Bouchaud.
\newblock Bubbles, crashes and intermittency in agent based market models.
\newblock {\em The European Physical Journal B-Condensed Matter and Complex
  Systems}, 31(3):421--437, 2003.

\bibitem{gomes2014market}
C.~Gomes and H.~Waelbroeck.
\newblock Is market impact a measure of the information value of trades? market
  response to liquidity vs. informed metaorders.
\newblock {\em Quantitative Finance}, (ahead-of-print):1--21, 2014.

\bibitem{hommes2005coordination}
C.~Hommes, J.~Sonnemans, J.~Tuinstra, and H.~Van~de Velden.
\newblock Coordination of expectations in asset pricing experiments.
\newblock {\em Review of Financial Studies}, 18(3):955--980, 2005.

\bibitem{joulin2008stock}
A.~Joulin, A.and~Lefevre, D.~Grunberg, and J.-P. Bouchaud.
\newblock Stock price jumps: news and volume play a minor role.
\newblock {\em Arxiv}, 2008.
\newblock \url{http://arxiv.org/abs/0803.1769}.

\bibitem{kindleberger2011manias}
C.~Kindleberger and R.~Aliber.
\newblock {\em Manias, {P}anics and {C}rashes: a {H}istory of {F}inancial
  {C}rises}.
\newblock Palgrave Macmillan, 2011.

\bibitem{kyle2012large}
A.~S. Kyle and A.~A. Obizhaeva.
\newblock Large bets and stock market crashes.
\newblock In {\em AFA 2013 San Diego Meetings Paper}, 2012.

\bibitem{lillo2005key}
F.~Lillo and J.~D. Farmer.
\newblock The key role of liquidity fluctuations in determining large price
  changes.
\newblock {\em Fluctuation and Noise Letters}, 5(02):L209--L216, 2005.

\bibitem{mackay2012extraordinary}
C.~Mackay.
\newblock {\em Extraordinary {P}opular {D}elusions and the {M}adness of
  {C}rowds}.
\newblock Start Publishing LLC, 2012.

\bibitem{mastromatteo2014agent}
I.~Mastromatteo, B.~Toth, and J.-P. Bouchaud.
\newblock Agent-based models for latent liquidity and concave price impact.
\newblock {\em Physical Review E}, 89(4):042805, 2014.

\bibitem{mastromatteo2014anomalous}
I.~Mastromatteo, B.~Toth, and J.-P. Bouchaud.
\newblock Anomalous impact in reaction-diffusion models.
\newblock {\em Physical Review Letters}, 113:268701, 2014.

\bibitem{moro2009market}
E.~Moro, J.~Vicente, L.~Moyano, A.~Gerig, J.~D. Farmer, G.~Vaglica, F.~Lillo,
  and R.~Mantegna.
\newblock Market impact and trading profile of hidden orders in stock markets.
\newblock {\em Physical Review E}, 80(6):066102, 2009.

\bibitem{nakamoto2008bitcoin}
S.~Nakamoto.
\newblock Bitcoin: A peer-to-peer electronic cash system.
\newblock {\em Consulted}, 1(2012):28, 2008.

\bibitem{reinhart2009time}
C.~Reinhart and K.~Rogoff.
\newblock {\em This {T}ime is {D}ifferent: {E}ight {C}enturies of {F}inancial
  {F}olly}.
\newblock princeton university press, 2009.

\bibitem{sandaas2001adverse}
P.~Sand{\aa}s.
\newblock Adverse selection and competitive market making: Empirical evidence
  from a limit order market.
\newblock {\em Review of Financial Studies}, 14(3):705--734, 2001.

\bibitem{shiller2013sharing}
R.~J. Shiller.
\newblock Sharing nobel honors, and agreeing to disagree.
\newblock {\em New York Times}, 26, 2013.

\bibitem{smith1988bubbles}
V.~Smith, G.~Suchanek, and A.~Williams.
\newblock Bubbles, crashes, and endogenous expectations in experimental spot
  asset markets.
\newblock {\em Econometrica: Journal of the Econometric Society}, pages
  1119--1151, 1988.

\bibitem{sornette2009stock}
D.~Sornette.
\newblock {\em Why {S}tock {M}arkets {C}rash: {C}ritical {E}vents in {C}omplex
  {F}inancial {S}ystems}.
\newblock Princeton University Press, 2009.

\bibitem{taleb2010black}
N.~N. Taleb.
\newblock {\em The black swan: The impact of the highly improbable fragility}.
\newblock Random House, 2010.

\bibitem{Barra:1997}
N.~Torre and M.~Ferrari.
\newblock Market impact model handbook, {BARRA} inc., {B}erkeley.
\newblock \url{http://www.barra.com/newsletter/nl168/mim4-168.asp}, (1997).

\bibitem{toth}
B.~Toth, Y.~Lempérière, C.~Deremble, J.~de~Lataillade, J.~Kockelkoren, and
  J.-P. Bouchaud.
\newblock Anomalous price impact and the critical nature of liquidity in
  financial markets.
\newblock {\em Phys Rev X}, 1:021006, 2011.

\bibitem{weber2005order}
P.~Weber and B.~Rosenow.
\newblock Order book approach to price impact.
\newblock {\em Quantitative Finance}, 5(4):357--364, 2005.

\bibitem{wyart2008relation}
M.~Wyart, J.-P. Bouchaud, J.~Kockelkoren, M.~Potters, and M.~Vettorazzo.
\newblock Relation between bid--ask spread, impact and volatility in
  order-driven markets.
\newblock {\em Quantitative Finance}, 8(1):41--57, 2008.

\end{thebibliography}

\end{document}